\documentclass[12pt]{article}

\usepackage{amsmath}
\usepackage{graphicx}
\usepackage{url} 
\usepackage{natbib}
\usepackage{xcolor}
\usepackage{amsmath}
\usepackage{amsfonts}
\usepackage{amssymb}
\usepackage{float}
\usepackage[inline]{enumitem}
\usepackage[nohyperlinks, printonlyused, smaller, nolist]{acronym}
\usepackage{makecell}
\usepackage{multirow}
\usepackage{algorithm}
\usepackage{algpseudocode}

\usepackage[hidelinks]{hyperref}
\usepackage{orcidlink}
\newcommand{\blind}{0}

\begin{document}

\def\spacingset#1{\renewcommand{\baselinestretch}%
{#1}\small\normalsize} \spacingset{1}


\if0\blind
{
  \title{\bf Uncertainty assessment of spatial dynamic microsimulations}
  \author{
    Morgane Dumont$^{1}$\thanks{The author gratefully acknowledges the financial support of the Fonds de la Recherche Scientifique – FNRS (Belgium) for funding the research stay of Morgane Dumont at Trier University.}~\orcidlink{0000-0003-2202-1542}, \\
    Ahmed Alsaloum$^{2}$\thanks{We are grateful to the German Research Foundation (DFG) for funding our research within the project DFG FOR 2559 MikroSim.}~\orcidlink{0009-0000-2121-9907}, \\
    Julian Ernst$^{2\dagger}$~\orcidlink{0009-0008-3679-1488}, \\
    Jan Weymeirsch$^{2\dagger}$~\orcidlink{0009-0004-6860-0448}, \\
    Ralf Münnich$^{2\dagger}$~\orcidlink{0000-0001-8285-5667} \\[0.5em]
    $^{1}$ HEC Liege - Management School of the University of Liège\\
    $^{2}$ Economic and Social Statistics Department, Trier University
}
  \maketitle
} \fi

\if1\blind
{
  \bigskip
  \bigskip
  \bigskip
  \begin{center}
    {\LARGE\bf Uncertainty assessment of spatial dynamic microsimulations}
  \end{center}
  \medskip
} \fi

\bigskip

\begin{abstract}
Spatial dynamic microsimulations probabilistically project geographically referenced units with individual characteristics over time. Like any projection method, their outcomes are inherently uncertain and sensitive to multiple factors. However, such factors are rarely addressed. Applying variance-based sensitivity analysis to both direct and indirect effects within the employment module of the MikroSim model for Germany, we show that commonly considered sources of uncertainty, namely coefficient and parameter uncertainty, are less influential than qualitative modeling choices. Because dynamic microsimulations are inherently complex and are computationally intensive, it is crucial to consider potential factors of uncertainty and their influence on simulation outputs in order to more carefully design simulation setups and better communicate results. We find, that simple summary measures insufficiently capture overall model uncertainty and urge modelers to account for these broader sources when designing microsimulations and their results.
\end{abstract}

\noindent%
{\it Keywords:}  sensitivity analysis, model evaluation, monte carlo simulation\\

\noindent%
Word count: 4994

\newpage
\spacingset{1.75} 

\section{Introduction}
\label{sec:intro}

Spatial dynamic microsimulations simulate the temporal evolution of populations at the individual (micro) level across different geographic zones~\citep{lovelaceDumontspatial2016}. In contrast to macrosimulation, which models population expectations, individuals are specifically represented and tracked over time. Due to an increase in computational power that enables more complex simulations, microsimulations have gained increasing prominence in both academic research (e.g. ~\citealp{bronka2025simpaths}) and official statistical institutions, such as Statistics Canada~\citep{ModGen2011} and Statistics Austria~\citep{statsim}. 

Within dynamic microsimulation, two types can be distinguished based on how time is treated. In continuous-time models, the duration until the occurrence of events is modeled, allowing analyses of exact timings of changes. In contrast, discrete-time microsimulations model the state of units at fixed, often yearly, intervals \citep{spielauer2009microsimulation}. This paper focuses on a discrete-time simulation incorporating household structures, where events affecting one individual can influence others within the same household or beyond. Microsimulations are helpful to assess the impacts of what-if scenarios, such as changes in policy \citep{li2013}. Depending on the research focus, various events or status transitions are incorporated into the simulation. In addition to randomness induced by the stochastic simulation approach, many other sources of uncertainty, such as modeling choices or parameter uncertainty, are relevant \citep{van1998microsimulation}.

Few articles detail the uncertainty in microsimulations, and the importance of its components is often overlooked. This is due to the high computational power required for such an endeavor, as a large number of different simulation configurations and individual runs are necessary for the analysis. Since microsimulations are typically developed to address specific problems and vary significantly in design and complexity, one-size-fits-all solutions are unattainable. Furthermore, there is no commonly agreed-upon standard for reporting uncertainty in microsimulations \citep{rissanen2024}.

While in simpler, static models, classical variance estimation methods like linearization or resampling can be applied to assess uncertainty (e.g., \citet{lappo} for sampling variability of the base data set), this is not the case for complex, dynamic models. \citet{sharif2012uncertainty} provides a guideline for applying an \ac{MC} approach to quantify uncertainty arising from parameter estimation by sampling from the parameters in dynamic models. Alternatively, sensitivity analyses can be applied to quantify how microsimulation outputs are affected by changes in the input factors \citep{schmausburgard_sensi}. This enables the inclusion of qualitative decisions and assumptions, in addition to the distribution-based uncertainty of parameters, in the quantification of overall uncertainty.

A better understanding of modeling uncertainties is necessary for designing models and interpreting their outputs. We aim to quantify the contributions of different uncertainty sources to the overall uncertainty using variance-based \ac{SA}. This may also contribute to an understanding of simulation design, particularly in light of the computational burden, to reduce unnecessarily large numbers of repetition due to diminishing returns on depicting overall uncertainty.

Our approach may be helpful as a blueprint for evaluating other microsimulation models in terms of sensitivity and variability. In this paper, the MikroSim model for Germany \citep{munnich2021population} is used to analyze uncertainty beyond the more often accounted for \ac{MC} effects. MikroSim is a suitable example, since it follows a discrete-time structure common to many dynamic microsimulations performed for different countries.

The paper is structured as follows: First, the underlying simulation model, uncertainty sources in microsimulation models, and the concept of variance-based \ac{SA} are explained. After, the simulation setup is described in Section~\ref{sec:simulationsetup}. The results are discussed in Section~\ref{sec:results}. Lastly, findings are summarized in Section~\ref{sec:conc}.

\section{Methods}
\subsection{The MikroSim model}
\label{sec:mikrosim}

MikroSim is a spatial dynamic microsimulation that projects a fully synthetic, geocoded representation of over 80 million individuals and over 40 million households in Germany. The base population is constructed from official statistical sources, including anonymized registers and the 2011 German census, as well as large-scale survey data. MikroSim is organized in a modular form within 14 transition modules \citep{munnich2021population}, each updating one thematically grouped set of variables. The modules are executed in sequential order, with recursive conditioning on previously simulated values to account for the stochastic dependence of previous transitions within the simulated interval \citep{galler1995competing}. Each module covers one aspect of life, including births, migration, deaths, household formations or separations, education, employment, and income \citep{munnich2021population}. 

\begin{figure}
    \centering
    \begin{minipage}{0.49\textwidth}
        \centering
        \includegraphics[width=0.8\textwidth,page=12]{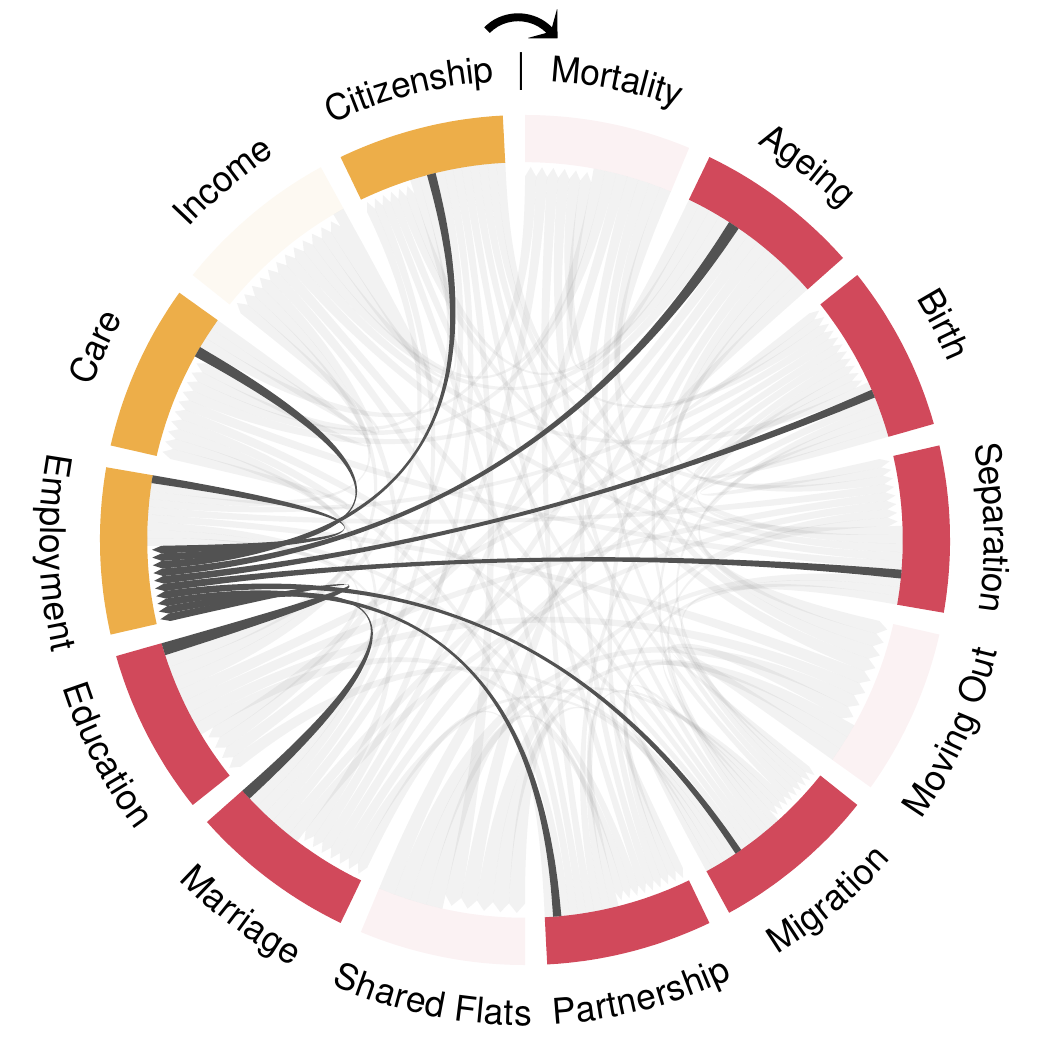} \\[-1em]
        \small a) Effects of other modules on employment
    \end{minipage}
    \begin{minipage}{0.49\textwidth}
        \centering
        \includegraphics[width=0.8\textwidth,page=12]{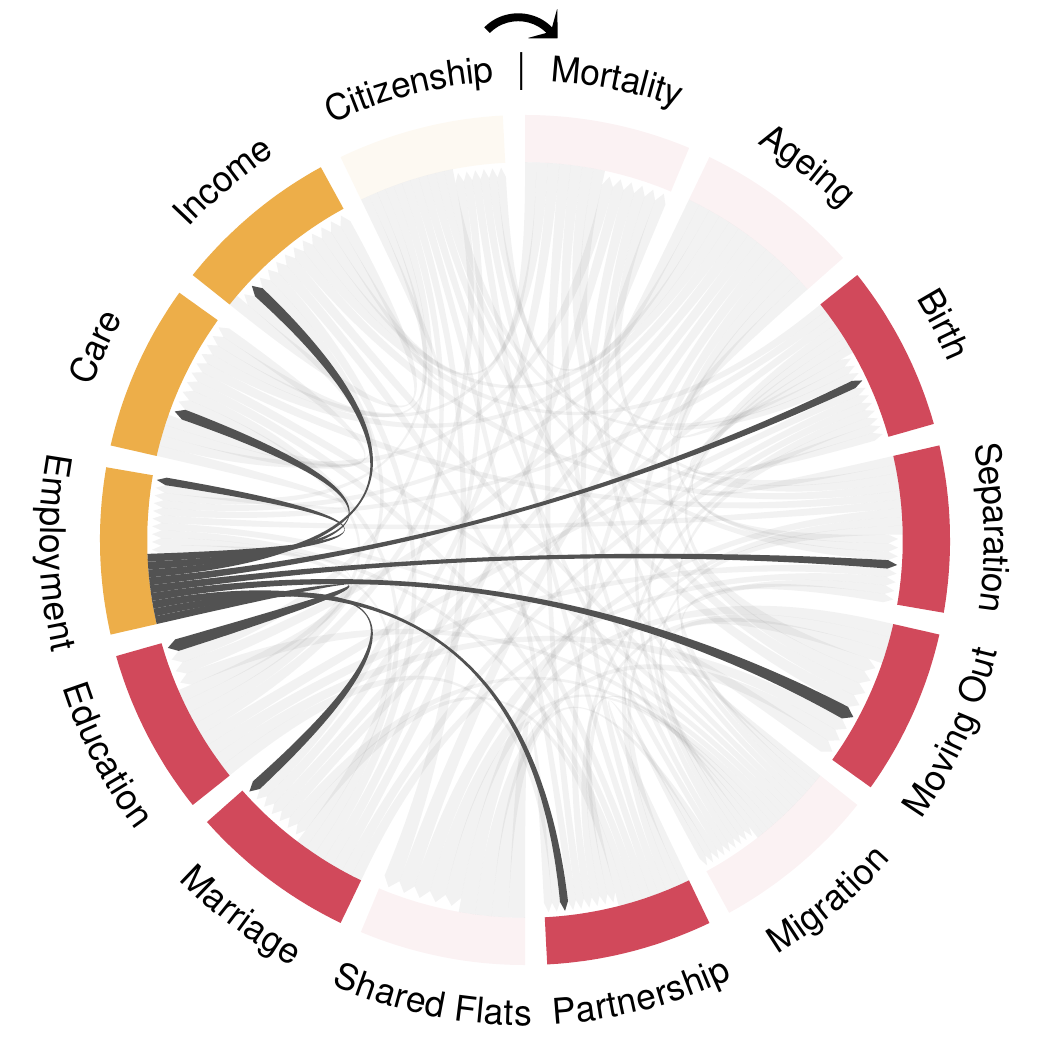} \\[-1em]
        \small b) Effects of employment on other modules
    \end{minipage}
    
    \vspace{0.5em}
    \definecolor{modo_next}{HTML}{EDAE49}
    \definecolor{modo_prev}{HTML}{D1495B}
    \definecolor{modo_arro}{HTML}{525252}
    \footnotesize{
         \textcolor{modo_prev}{$\blacksquare$}~Last update in $t$ \quad \textcolor{modo_next}{$\blacksquare$}~Last update in $t-1$ \quad \textcolor{modo_arro}{$\blacktriangleright$}~Direction of interaction
    }
    \caption{Module interactions with employment in the MikroSim framework.}
    \label{fig:module_interactions}
\end{figure}

Typically, status changes are modeled stochastically based on estimated transition models. With the extent, level of detail, and structure of the simulation model, it is computationally unfeasible to consider the uncertainty caused by each module individually. Thus, this paper focuses on the uncertainty of one particularly inter-correlated model: employment. The interdependency with other modules in the current and next simulation interval is illustrated in Figure~\ref{fig:module_interactions}. 
The employment module is especially well-suited for such analysis, as only the mortality and shared flats modules have no direct interactions with it. Note that module interactions may vary in kind and strength. These interactions may be in the form of a filter, e.g., individuals under 15 year are not eligible to work and thus deterministically inactive, or as a predictor in the model. At the same time, aging has a direct model effect on the transitions between employment groups. The strength of the effects depends on the model parameters or coefficients, which in turn may also be influenced by the model complexity, as well as other modeling factors.

Within the employment module, the labor participation is determined first, and working time for the employed is determined afterward. As the employment status is usually of more interest, the respective model will be the focus of this paper. Labor participation in MikroSim is modeled according to the definition of the International Labor Organization\footnote{\url{https://ilostat.ilo.org/methods/concepts-and-definitions/description-work-statistics-icls19/\#elementor-toc\_\_heading-anchor-3}}, followed by the German microcensus for the population of 15 to 74-year-old in three categories: employed, unemployed, and economically inactive, with yearly transition probabilities. 

Since the base population is rooted in 2011, transition probabilities within all relevant demographic modules are aligned towards known totals for the simulation years 2011~-- 2022 to ensure consistency, using the bi-proportional logit-scaling algorithm proposed by \citet{stephensen2016}. Based on this, modified intercept values are calculated for the model outputs \citep{schmaus2023, weymeirsch2024model} on the German district level, which may be applied after the calibration phase. This procedure, which \citet[][pp.~62]{schmaus2023} refers to as \emph{regionalization}, is required because transition models are typically estimated through surveys without granular geographic markers (often on the national level), thus insufficiently capturing regional differences in outputs.

For employment, census values from 2011 and 2022 are available at the district level. 
Derived adjustment values from this alignment period can also be used for the projection period. In particular, we follow the \emph{last-carried-forward} approach, where the adjustment values are used to modify the transition probabilities in the projection phase. 

\subsection{Uncertainty in dynamic microsimulations}
\label{sec:uncertainty}

In dynamic microsimulations, the characteristics of the units, like education level or partnership status, are either stochastically or deterministically updated as time passes.  
Due to the stochastic updating via \ac{MC} processes, simulation outcomes vary upon repetition of the simulation. In addition, many types of uncertainty are relevant to the simulation outcomes \citep{schmaus2023,sharif2012uncertainty,van1998microsimulation,bilcke2011accounting}. 

When comparing simulation results, it is crucial to determine whether differences are due to procedural randomness or systematic effects. However, as noted by \citet{goedeme2013testing}, this is often not considered. Further, there are no commonly agreed-upon standards for dealing with and reporting of uncertainty in the field \citep{rissanen2024}. Since dynamic microsimulation models tend towards high complexity and are characterized by variable interactions and feedback effects, assessing uncertainty is not straightforward. 
Different types of uncertainty are relevant:

\textbf{Monte Carlo uncertainty:} The \ac{MC} uncertainty, also referred to as 'model-specific' uncertainty \citep{Jia2023Moving}, is caused by stochastic procedures within a simulation. One possible approach to assessing this uncertainty is to determine the total variation of the output by repeating the simulation multiple times and summarizing the results afterward using uncertainty intervals. 

\textbf{Methodological uncertainty:} Many methodological decisions induce uncertainty even before running a microsimulation, since a different choice would lead to different outputs. First, the choice of the model (type and complexity), which determine the fit to the data as well as its applicability for projections, could have substantial effects on the simulation outcome. This is especially the case for dynamic microsimulations, where outputs of one module are used for the stochastic process of another. Increasing the complexity of the model implies estimating additional parameters, thus eventually increasing the variability. As a remedy, model complexity should stay appropriately low \citep{van1998microsimulation, statsim}. Secondly, a starting population is required. Since, in practice, no data set with all necessary variables exists for the whole population, the starting units are often at least partially synthetic~\citep{rahman2016small,Gallagher2018}. Repeating the population synthesis process can lead to variations when using non-deterministic methods, and methodological choices and assumptions may induce further uncertainty. This is linked to the \ac{MC} uncertainty of the population synthesis used as input for the dynamic microsimulation. In addition, the structure and coordination of the different models also induce uncertainty. Indeed, if transition probabilities for discrete-time models cannot be conditioned properly on previous modules, the order of modules becomes relevant due to competing events, especially when time-steps are large \citep{van1998microsimulation,galler1995competing}. In such instances, simulation outcomes are sensitive to the order, as it determines whether a unit is further simulated or not \citep{dumont2018}.

\textbf{Parameter uncertainty:} The specification randomness of stochastic models arises due to the fact that model coefficients must be estimated, e.g., from surveys, such that sampling and non-sampling errors become relevant \citep{pudney1994reliable, van1998microsimulation}. This includes the choice of data to estimate from, the treatment of missing values, and the decisions regarding imputation. To analyze the uncertainty caused by the estimation of input parameters, \ac{SA} can be conducted. Hereby, the most influential input factors are identified to determine the extent to which they contribute to the overall uncertainty of the model results. This helps to better understand model behavior and complexity \citep{Janssen2012methodsSA}.

\textbf{Assumption and scenario uncertainty:} Microsimulations are often used to simulate the impact of new policies. The way these existing or new policies and baseline scenarios are translated in to simulation configurations may impact results. Such political scenarios are not considered here. However, in the literature, "scenario" is also used to refer to exogenous assumptions used as input for the model. In this paper, we distinguish the uncertainty caused by political scenarios and assumption. Indeed, for some modules, transitions are inherently difficult to estimate or calibrate with historical data only. For example, immigration is driven mostly by external factors not specifically simulated, like war or natural catastrophes in other countries. Further assumptions about the future development, like the future development of life expectancy, are also inherently uncertain. To reflect uncertainty about such assumptions, multiple variants of assumptions are usually considered, e.g., low or high migration scenarios \citep{o2001guide}.

\subsection{Variance-based sensitivity analysis}
\label{sec:uncertain_measures}

In complex simulations involving both quantitative and qualitative inputs, assessing uncertainty can be challenging. Variance-based \ac{SA} can be used to analyze the first-order effect (main effect) and the total effect of the individual input variable, including interaction effects with other variables, by decomposing the total variance into conditional variances. It is particularly suitable for microsimulation models to measure the contributions of the individual inputs to the overall uncertainty of the model, as it does not require a large sample from the input uncertainty range, which, especially in microsimulation models, often includes qualitative assumptions.

We consider the microsimulation model as a function of structures and hypotheses, which can be parameterized or non-parameterized as inputs. Since the model is a dynamic system, we take the time period into account. Thus, the target value of the model is a function of input variables as follows:
$Y^{(s)}=f(X_1,X_2,\cdots ,X_k,s)$, where $s$ is the simulation period and $(X_1,X_2,\cdots, X_n)$  are the k-dimensional pairwise independent input factors, which are considered as different scenarios, models, parameters, or data sources.

Then the variance decomposition for $\text{Var}(Y^{(s)})$ is given by:
\begin{flalign}
    \label{eq:vardecomp}
    \text{Var}(Y^{(s)}) &= \sum_{i=1}^{k} V_i^{(s)} + \sum_{i=1}^{k} \sum_{j=i+1}^{k} V_{ij}^{(s)} + \ldots + V_{1,\ldots,k}^{(s)} &&\\
    \label{eq:vc}
    \text{with \qquad\qquad}
    V_i^{(s)} &= \text{Var} \left\{ \mathbb{E}[ f (X) | X_i ] \right\} &&\\
    \label{eq:vcint}
    V_{ij}^{(s)} &= \text{Var} \left\{ \mathbb{E}[ f (X) | X_i, X_j ] \right\} - V_i^{(s)} - V_j^{(s)} &&\\
    V_{i_1 \cdots i_d}^{(s)} &= \text{Var} \left\{ \mathbb{E}[ f (X) | X_{i_1}, \ldots, X_{i_d} ] \right\} - \sum_{m=1}^{d-1} \sum_{\substack{j_1, \ldots, j_m \\ \in (i_1, \ldots, i_d)}} V_{j_1 \cdots j_m}^{(s)} &&
\end{flalign}

where $ \mathbb{E}[ f (X) | X_i ] $ is the expected value of $f$ given $X_i$
. Consequently, Equation~\ref{eq:vc} denotes the variance of the expected value of $f$ if only $X_i$ is varied. $V_i^{(s)}$ is called the \ac{VC} or uncertainty component of the factor $X_i$ in simulation period $s$ \citep{schmaus2023,schmausburgard_sensi}.

The proportion of variance, or \ac{SI}, caused by a single input variable $X_i$ alone is called the first-order or \ac{mSI} $S_i$, which is computed as follows \citep{Saltelli2008}: 
\begin{equation}
    S_i^{(s)} = \frac{V_i^{(s)}}{V(Y)^{(s)}} = \frac{\text{Var}[\mathbb{E}(Y|X_i)]}{\text{Var}(Y)}\,
\end{equation}

Furthermore, the proportion of the two input variables $X_i$ and $X_j$ is called the second-order or \ac{iSI}, which is given by \citep{Saltelli2008}:
\begin{equation}
    S_{ij}^{(s)} = \frac{V_{ij}^{(s)}}{V(Y)^{(s)}} = \frac{\text{Var}[\mathbb{E}(Y|X_i,X_j)] -V_{i}^{(s)} - V_{j}^{(s)}}{\text{Var}(Y)}\, 
\end{equation}

To capture the overall sensitivity of an input variable  $X_i$, i.e., the main effect and all interaction effects of $X_i$, the \ac{tSI} $S_{Ti}$ is calculated for factor $i$ :
\begin{equation} 
    S_{Ti}^{(s)} = \frac{\text{Var}(Y) - \text{Var}(\mathbb{E}[Y | X_{\sim i}])}{\text{Var}(Y)}
\end{equation}

where  $X_{\sim i} $   denotes all factors except $X_i$. The \acp{tSI} and \acp{VC} are used to measure the minimum influence of each factor.

\section{Simulation setup}
\label{sec:simulationsetup}

To quantify the impact of uncertainty sources on the \ac{SI} or \ac{VC} measures, the MikroSim model is run with varying configurations related to the uncertainty drivers outlined in Section \ref{sec:uncertainty}. To reduce the computational burden, the simulation and analysis are limited to the 36 districts in the federal state of Rhineland-Palatinate, Germany, which encompasses both very small (Zweibrücken: 34,000 inhabitants) and large (Mainz: 220,000 inhabitants) districts with varying degrees of urbanization. To assess the influence of specific configurations on the employment modeling process, other modules and simulation parameters not part of this analysis are kept constant. Model outcomes are aligned to observed regional values within the benchmark period (2011 -- 2022) across all modules where historic values are available\footnote{In detail, these are: Mortality, Birth, Migration, Marriage, Education, Employment, Care, Citizenship.} in order to reduce variance and meet expectations from observed reality. Intercept adjustment terms, derived from the alignment procedure, are then used for the projection phase (2023 onward) within all affected modules. 

\subsection{Methodological uncertainty} 

One contributing factor to methodological uncertainty is the choice of the modeling approach. Typically, regression models are applied to estimate state transitions in microsimulations. However, machine learning methods could also be considered. The type of model may have an effect, as it influences the model's fit to the data and therefore determines granular aspects of the predictions. To capture this aspect of modeling choice, two \emph{model types}, \ac{MNL} and \ac{RF}, are considered. 

Apart from model type, model specifications, including the covariates included, influence predicted outcomes. Three \emph{model complexities} are considered. All models were estimated separately by sex and are fully interacted with the employment status of the previous period $t-1$, in order to allow varying effects of the covariates depending on the current state. The low complexity model only accounts for age and previous employment. In the medium complexity model, person-level variables on birth events and immigration in the same simulation period, and information on educational level, care status, and citizenship are added. Finally, in the highest complexity model, information on partnership and household, namely the marriage status (single, partnered with cohabitation, and married with cohabitation), and the number of children, and the age of the youngest child are included. Note that for all \ac{MNL} models, splines are used to account for non-linearities in age and duration-related variables (age, years since immigration, and age of the youngest child), and squared terms for the \ac{RF} models\footnote{All models have high explanatory power of around 58\% deviance explained for the low and 62\% for the high complexity models. The out-of-sample Brier score is improving with complexity, with the \ac{RF} having a slightly better performance than \ac{MNL} in all settings.}. An overview of the variables in the model complexities is given in Table \ref{tab:modelvars}.

\begin{table}[htb]
    \newcommand{\x}[0]{{$\times$}}
    \centering
    \begin{tabular}{l|c|c|c}
        \hline\hline
        \textbf{Variables} & \multicolumn{3}{c}{\textbf{Complexity}} \\
        \cline{2-4}        & Low & Medium & High \\ \hline
        employment$^{t-1}$ & \x  & \x & \x \\ \hline
        age   & \x  & \x & \x \\ \hline
        birth event        & { } & \x & \x \\ \hline
        citizenship$^{t-1}$ & { } & \x & \x \\ \hline
        years since immigration & { } & \x & \x \\ \hline
        education level      & { } & \x & \x \\ \hline
        care status$^{t-1}$  & { } & \x & \x \\ \hline
        partnership \& marital status   & { } & { } & \x \\ \hline
        number of children in HH  & { } & { } & \x \\ \hline
        age youngest child  & { } & { } & \x \\ \hline
        \hline
    \end{tabular}
    \caption{Predictor variables for employment categories, separately estimated by sex}
    \label{tab:modelvars}
\end{table}

A further decision needs to be made on whether to regionalize the models beyond aligning the transitions for the period for which observed values are available. More precisely, whether the derived \emph{intercept adjustment} values in the alignment period (2011 -- 2022) for the employment module are used to adjust the projection into the future (2023 onward) or if the simulation should rely solely on the model probabilities. Both possible decisions are considered. We expect this factor to have a less significant impact in regions that are structurally similar to the microcensus population and a higher impact in regions that either inherit unobserved differences or are otherwise structurally divergent. We further expect a strong interaction between adjustment and model complexity, as more complex models capture more of the regional heterogeneity and are thus less reliant on intercept correction to reproduce regionally observed totals. 

\subsection{Parameter uncertainty} 

The parameters of all twelve models (two \emph{model types}, three \emph{model complexities}, each for male \& female) are estimated and thus inherently uncertain. Further randomness is introduced by choices regarding the data source, such as the dataset or the years used for estimation. To account for this, two waves of the scientific-use file of the German microcensus panel, a compulsory $0.7\%$ household sample of Germany, are used (2012~--~2015 and 2016~--~2019). While cross-sectional missingness is almost non-existent, previous-year information ($t-1$) that cannot be deterministically derived from the cross-sectional information must be imputed for individuals moving into dwellings selected for the panel. To reflect the additional uncertainty introduced by the imputation, the estimated \ac{MNL} produced for each of the $m = 5$ imputed data sets was pooled according to the combination rules of \citet{Rubin1987}, while the predicted probabilities of the \ac{RF} were averaged across the respective models. A more thorough description of the data set, along with an equivalent imputation procedure, can be found in \citet{weymeirsch2024model}.

Like \citet{bronka2025simpaths}, \citet{sharif2012uncertainty} and \citet{petrik2020uncertainty}, we sample parameter estimates by drawing from a multivariate normal distribution using the estimated coefficients as mean and covariances of the model coefficients to account for parameter uncertainties. Note that this was not done for the \ac{RF} where no coefficients exist. Thus, when evaluating both the \ac{RF} and \ac{MNL} jointly in Section~\ref{sec:results}, only the \ac{MNL} point estimates (\emph{coefficient draw}~0) are considered.

\subsection{Monte Carlo uncertainty} 

The \ac{MC} uncertainty is considered by running each configuration several times. 
We split \ac{MC} effects from stochastic processes in the employment module (\emph{intmc}) from those in other modules (\emph{extmc}), by using two sets of random number seeds within each simulation run.

The external \ac{MC} factor, used for all stochastic processes in our simulation outside the employment module, varies for each \emph{district}~$\times$ \emph{simulation year} factor combination, as well as its own \emph{static component}. This allows us to keep stochastic processes in the microsimulation relatively across all other analysis factors, as the seed only varies for the information within the compound value. We can therefore attribute effects caused by the \emph{extmc} factor directly to \ac{MC} effects caused in other modules or on rebound effects of the simulation factors on employment.

The internal \ac{MC} factor, which is only used within the employment module, varies for each \emph{district}~$\times$ \emph{simulation year}~$\times$ \emph{extmc} combination, as well as its own \emph{static component}, meaning that it is fixed for any of the other factors (model type, model complexity, model period, migration scenario, intercept adjustment, coefficient draw; see Table~\ref{tab:scenarios} on page~\pageref{tab:scenarios}). Any effects on sensitivity caused by the \emph{intmc} factor can therefore be attributed to the primary, secondary, and rebound \ac{MC} variation of the employment module itself.

\subsection{Assumption uncertainty} 

For any projection, assumptions about future evolution are made, which are inherently uncertain. For demographic development, this uncertainty is typically reflected by specifying different scenarios for the demographic change components \citep{o2001guide}. Especially on small geographic scales, migration is the single most important component of population change and at the same time the most difficult aspect to forecast \citep{wilson2022preparing, wilson2005recent}, with distinct and spatially diverse impacts on simulation outcomes \citep[][pp.~139]{ernst2023influence, schmaus2023}. The uncertainty about the overall level and composition of migration thus needs special consideration in the simulation. 
Until 2021, regional migration data by movement type, age-group, sex and citizenship type are integrated into the simulation to replicate observed developments \citep[][pp.~147]{schmaus2023}.

After 2021, two projection migration scenarios are considered. First, (\emph{Full}), the averaged total and distribution of internal and external migration observed from 2011 to 2021 for each district by age, sex, and citizenship group are assumed for future years. For the second migration scenario (\emph{Selected}), selected years are excluded from the calculation of the average. Namely, the year 2020, which saw reduced migration due to the SARS-CoV-2 pandemic, and 2015~--~2016, in which larger immigration flows occurred for Germany in the wake of the Syrian civil war. Thus, the scenarios vary both in overall number and composition of migrants. We expect the spatial patterns of these differences to be relatively diverse across the entire simulated geographical region. For a more detailed description on migration module scenarios in MikroSim, we refer to \citet{ernst2023influence}.

\subsection{Summary of Simulation Configuration}
\label{sec:simsetupsummary}

\begin{figure}[!b]
    \centering
    \includegraphics[width=0.73\textwidth]{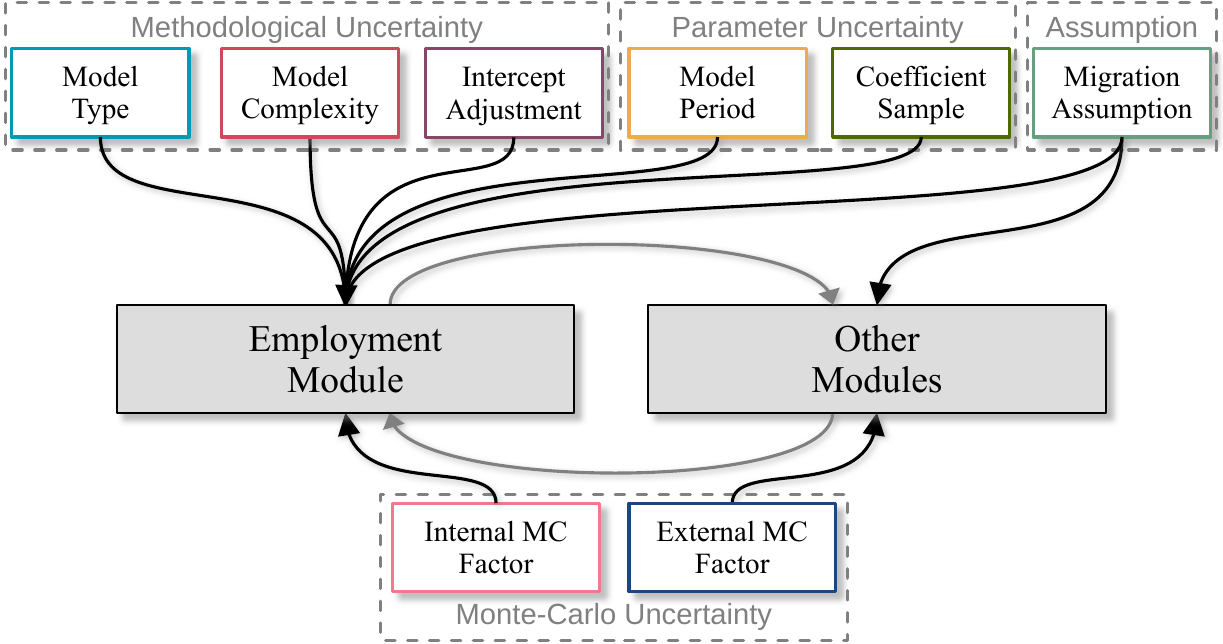}
    \caption{Influences of uncertainty types on simulation modules}
    \label{fig:factorinfluence}
\end{figure}

A schematic of factors influencing the employment and other modules, as well as their interaction is depicted in Figure~\ref{fig:factorinfluence}. In our simulation setup, several factors directly influence the employment module, which is the main target of our analysis. The migration scenario has a direct influence on almost all modules, including employment, as it changes the composition of the entire population. Contrary to that, the external \ac{MC} factor influences all modules that use stochastic processes, except for the employment module, which instead is influenced by the internal \ac{MC} factor. Any changes to the population have secondary effects, due to changed characteristics of individuals and the high degree of interlinking modules (refer to Figure~\ref{fig:module_interactions}). Other modules in turn can affect the employment module in the next iteration. 

\begin{table}[htb]
    \begin{tabular}{l|l|r|l}
        \hline\hline
        \textbf{Uncertainty} & \textbf{Factor Name} & \textbf{Dim.}  & \textbf{Level Description} \\ \hline
        \multirow{3}{*}{Method} &
                      Model type (\texttt{type})           & 2  & MNL, RF \\
        \cline{2-4} & Model complexity (\texttt{compl})    & 3 & Low, Medium, High \\
        \cline{2-4} & Intercept Adjustment (\texttt{adju}) & 2  & Yes, No \\
        \hline
        \multirow{2}{*}{Parameter} &
                       Model period (\texttt{period})          & 2  & 2012--2015, 2016--2019 \\
         \cline{2-4} & Coefficient draw$^{*}$ (\texttt{coeff}) & 11 & 0$^{**}$, 1~--~10 \\
         \hline
        \multirow{2}{*}{Monte-Carlo} & External \ac{MC} (\texttt{extmc})  & 5  & 1~--~5 \\
        \cline{2-4} & Internal \ac{MC} (\texttt{intmc})  & 10 & 1~--~10 \\
         \hline
        Assumption & Migration scenario (\texttt{migr})     & 2  & Full, Selected\\
        \hline\hline
    \end{tabular}
    \newline
    \begin{footnotesize}
        \textit{%
        \\[-1em]
            *) Only for \texttt{MNL}, whereas just a single dimension in \texttt{RF}.\\[-1em]
            **) No coefficient draws were taken in the '\texttt{coeff} = 0' scenario, and instead point estimates predicted.
        }
    \end{footnotesize}
    \caption{Simulation scenarios and sensitivity factors}
    \label{tab:scenarios}
\end{table}

Table~\ref{tab:scenarios} summarizes the different factor combinations and their resulting dimensions. To fully assess the sensitivity caused by each factor, every factor combination is run separately. As coefficient uncertainty is only considered for the \ac{MNL}, this leads to 13,200 combinations for the \ac{MNL} and 1,200 for the \ac{RF}. Thus, across 36 districts, this results in a total of 518,000 runs. Sensitivity was calculated for a total of ten directly and indirectly affected indicators, namely the overall unemployment rate, unemployment rates by citizenship and sex, the share of mothers in employment, the share in partnerships, the total fertility rate, the average household size and the share of population under 18.

\section{Results and discussion}
\label{sec:results}

The $518,000$ simulation runs are computed on a parallelization infrastructure simultaneously across multiple thousand computation cores. On average, each individual simulation run took one hour and twenty minutes, using $2.5$ GB of memory at peak during the process, underscoring the heavy computational burden required for a \ac{SA} even when considering only a single module of a complex dynamic microsimulation.

\begin{table}[htb]
    \centering
\begin{tabular}{ll|llllllll}
   \hline\hline
   \textbf{Factor} & { } & type & compl & adju & period & coeff & extmc & intmc & migr \\
   \hline
   \multirow{2}{*}{\textbf{\makecell{Incl. low\\[-1em]complexity}}} & Max. & 0.061 & 0.893 & 0.385 & 0.137 & 0.007 & 0.024 & 0.285 & 0.926 \\
    { } & Median & 0.016 & 0.464 & 0.048 & 0.041 & 0.002 & 0.000 & 0.021 & 0.001 \\
    \hline
   \multirow{2}{*}{\textbf{\makecell{Excl. low\\[-1em]complexity}}} & Max. & 0.199 & 0.169 & 0.559 & 0.300 & 0.031 & 0.055 & 0.528 & 0.927 \\
   { } & Median & 0.047 & 0.031 & 0.219 & 0.104 & 0.004 & 0.002 & 0.041 & 0.001 \\
   \hline\hline
\end{tabular}
    \caption{Summary measures of first-order sensitivity indices in 2040 by factor}
    \label{tab:max_std_SI_2040}
\end{table}

The maximum \ac{mSI} across all outputs for the contributing factors is shown in Table~\ref{tab:max_std_SI_2040}. Due to the stark differences caused by the low complexity models, the values within the table are divided into measures including or excluding them in the \emph{compl} factor.

With a maximum well below 1\% overall and around 3\% when low complexity models are excluded, the \emph{coeff} factor barely contributes to the variability in the simulations (median lower than 0.4\%) even on the district level. This is likely due to the large sample size of the survey panel, which leads to small standard errors. Because the \emph{type} and \emph{coeff} factors are mutually exclusive and would need separate analysis, we drop \emph{coeff} in the following.

Including the low-complexity model leads to a substantial dominance of the \emph{compl} factor, particularly for more direct output variables, such as unemployment rates, even when grouped by nationality or sex. As they only consider age and previous employment status, the low complexity models create strong variability in other population subgroups. While simpler models are more characteristic of macro approaches, microsimulations typically consider more individual or household information. We consider the low-complexity model unlikely to be used in general microsimulation applications. The analyses are therefore separated into two cases: with and without the low complexity models.

\begin{figure}[!htb] 
    \centering
    \includegraphics[width=\textwidth]{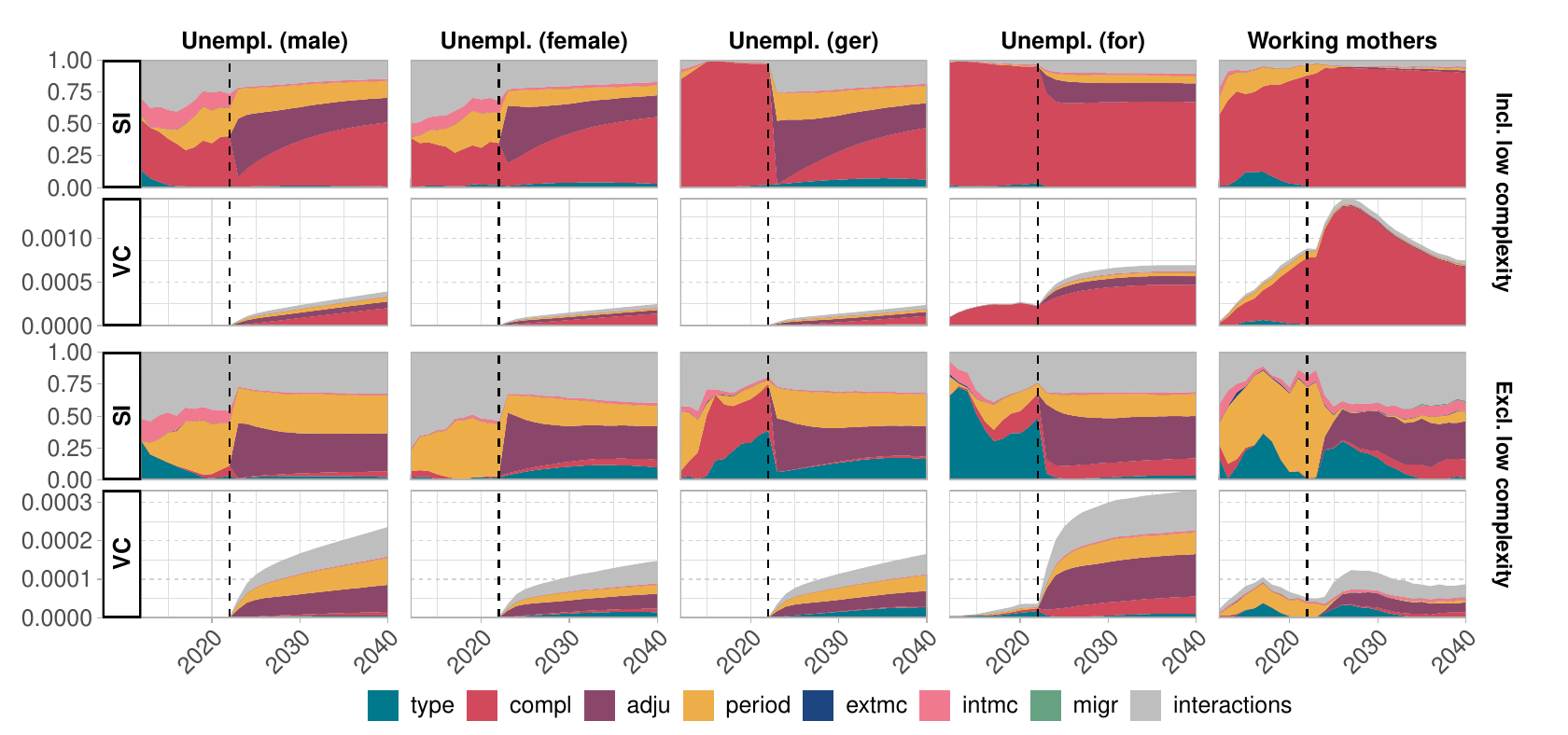}
    \caption{Sensitivity indices and variance components for selected direct indicators}
    \label{fig:area_chart}
\end{figure}

In our analysis, for most indicators, the \ac{VC} is strictly increasing with time (see Figure~\ref{fig:area_chart}). As only the overall labor participation categories are aligned until 2022, to replicate observed developments, there is low, but not zero, variation in subgroups such as unemployment by nationality or sex during this period. Notably, the working mothers indicator is already subject to considerable variability before the projection phase. For all indicators, there is a strong sensitivity towards including the low-complexity model, which dominates the index at any point in time. This is especially true for subgroups not specifically considered in the low-complexity model, like mothers or the decomposition by nationality. 

In the projection phase, the \ac{VC} is strongly increasing for most indicators. Apart from a slight reduction in sensitivity towards the decision for or against model adjustment across time, there is no notable change in pattern for the factors. This holds both at the federal level and at the district level. Thus, for simplicity, further analyses are conducted cross-sectionally for 2040 hereafter. 

\begin{figure}[htb] 
    \centering
    \includegraphics[width=0.99\textwidth]{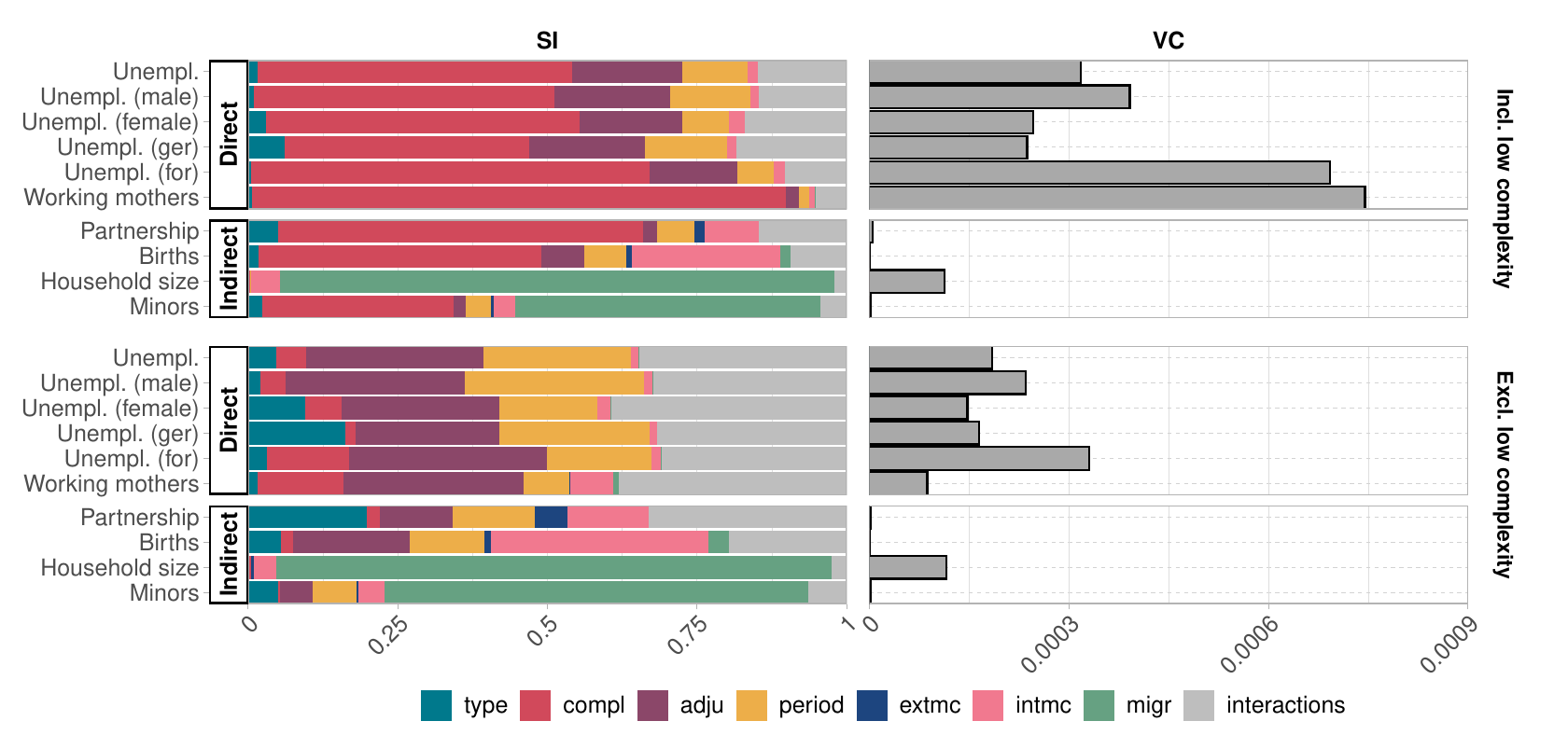}
    \caption{Sensitivity indices and variance components in 2040}
    \label{fig:SI2040_direct}
\end{figure}

Figure~\ref{fig:SI2040_direct} shows the cross-sectional sensitivity for the selected directly and indirectly influenced indicators by the \ac{mSI}, \ac{iSI}, and \ac{VC} for the year 2040 on the federal state level. As only the employment module is varied, direct indicators are more strongly influenced than indirectly affected outcomes. Regardless of the consideration of the low complexity model, the share of working mothers is the most sensitive among all indicators. 

As can be seen, the \ac{mSI} is indeed mostly driven by the inclusion of the low-complexity model. As expected, indirectly affected indicators are less sensitive to changes in the employment module, with some being almost invariant. While household sizes are more subject to variation than other indirect indicators, this is mostly a result of migration. Excluding the low-complexity model from the analysis, model outputs vary less overall, and sensitivity is now mostly influenced by decisions about whether model probabilities are regionalized towards observed values or not, and the model estimation period. On the federal state level, \ac{MC} uncertainty and coefficient uncertainty are not relevant factors for a full population model and models estimated on large surveys. 

\begin{figure}[htb] 
    \centering
    \includegraphics[width=0.99\textwidth]{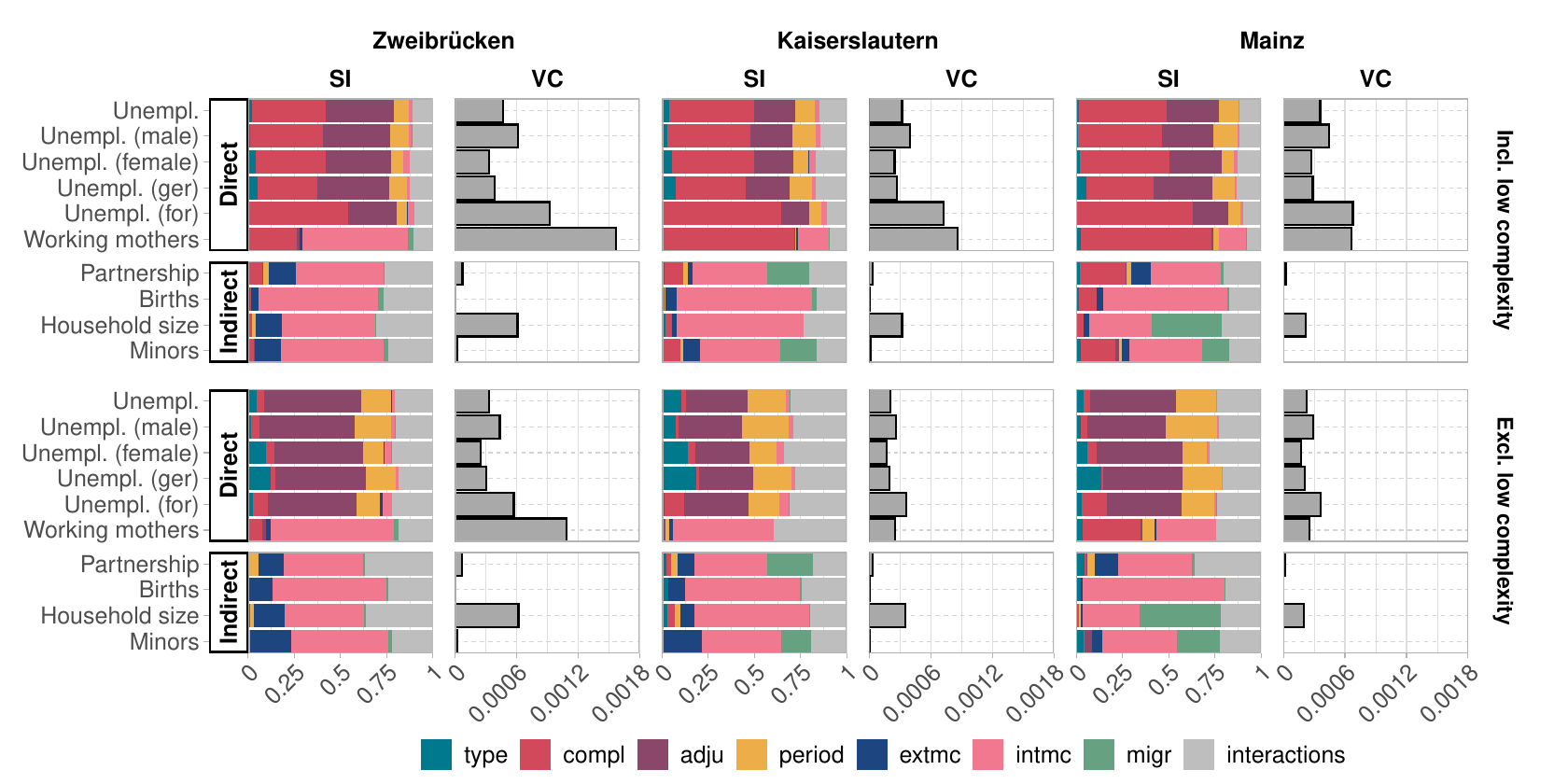}
    \caption{Sensitivity indices for selected districts in 2040}
    \label{fig:SI2040_direct_district}
\end{figure}

At district level, \ac{MC} is the predominant driver in indirect effects and a relevant factor for indicators in subgroups with smaller risk populations (see Figure~\ref{fig:SI2040_direct_district}). As expected, the effect of \ac{MC} decreases with population size and is less relevant in the district of Kaiserslautern (over 100,000 inhabitants) and Mainz (around 225,000 inhabitants) than Zweibrücken (just over 30,000 inhabitants). Note that even the latter is comparably large for most microsimulation models, which often only simulate a sample rather than the entire population. 

Decisions whether to post-adjust the model outputs via intercept adjustment to better fit locally observed shares and totals are, naturally, more influential on the regional level than on the federal state level, where effects may be averaged out to a degree. Once the low complexity models are excluded, model type becomes more relevant than complexity. Interestingly, this is not very relevant for the working mothers, where complex interactions likely better captured by the \ac{RF}, would be expected to be more influential. 

\begin{figure}[htb]
    \centering
    \includegraphics[width=\textwidth]{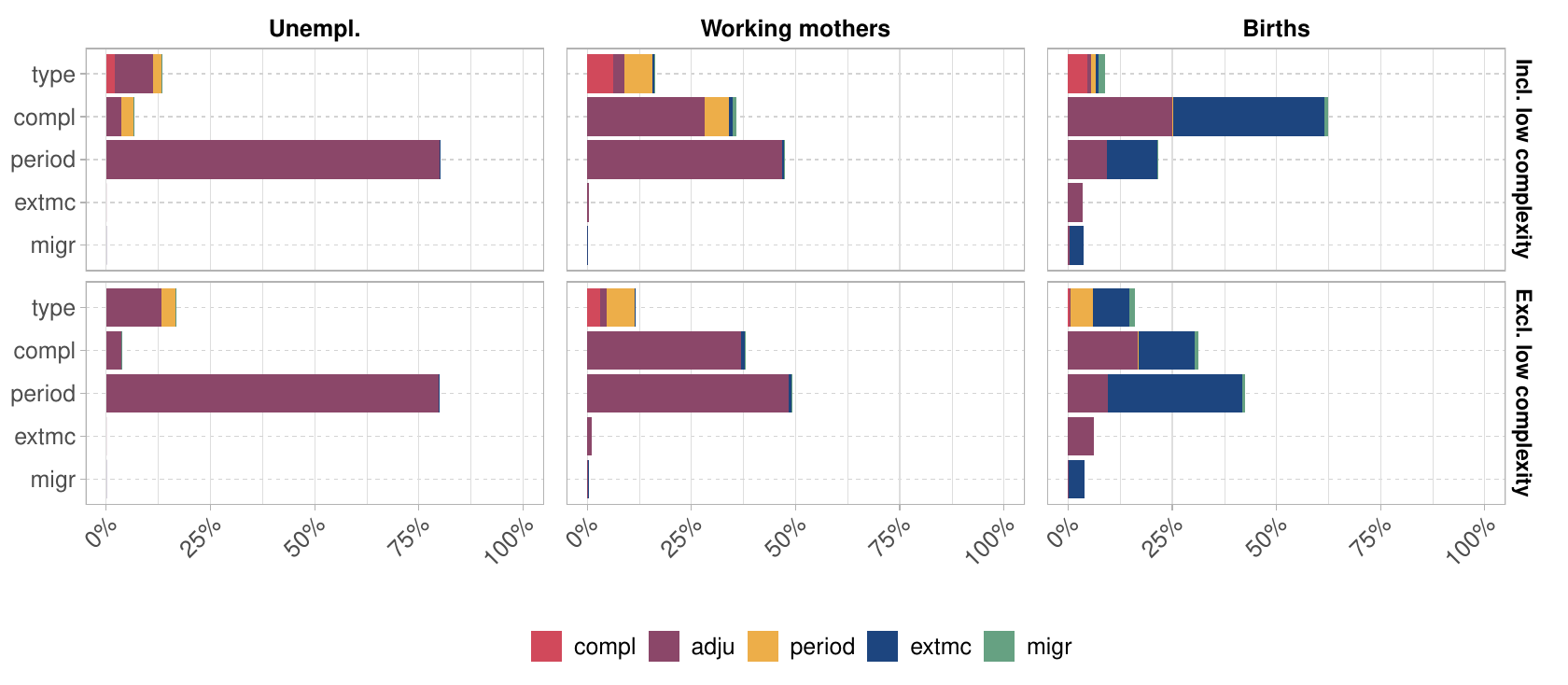}
    \caption{Breakdown of the interaction effect by factors}
    \label{fig:interaction}
\end{figure}

Interactions make up a large portion of the overall sensitivity at all points in time and are thus analyzed in more detail. As can be seen in Figure \ref{fig:interaction}, interactions of \textit{adju} make up most of the interaction effects for all indicators. For the direct indicators, the interaction with the model estimation period is the dominant factor. As models estimated on the two time periods have a different fit for the calibration phase, the severity of adjustment to match observed values differs, leading to a weaker or stronger effect of \textit{adju} depending on which model period is considered. For the subgroup indicator of working mothers, the interaction between the adjustment decision and model complexity becomes relevant. Interestingly, this holds regardless of the inclusion of the low-complexity model, which makes only a minor contribution to the interaction effects. For the indirect indicators which are directly influenced by models outside employment, the \ac{MC} effect via the \emph{extmc} factor is a large contributor to the effect of interactions.

\section{Conclusion}
\label{sec:conc}

In this paper, we analyzed the sensitivity of spatial dynamic microsimulation outputs exemplarily for the employment module of the MikroSim model for Germany. Methodological uncertainty, parameter uncertainty, \ac{MC} uncertainty, as well as uncertainty regarding qualitative assumptions were considered.

Overall, for full-population models, qualitative modeling decisions, such as calibrating the model or choosing the estimation data, appear to be more influential than the randomness produced within the simulation, which is the focus of most analyses. Notably, the more relevant factors are not reduced by increasing simulation runs, simulated population sizes or survey data size. Thus, a level of uncertainty is always present due to methodological choices. Simply considering \ac{MC} and parameter uncertainty by repeated simulations, which could then be easily summarized via traditional uncertainty intervals, clearly does not capture overall uncertainty sufficiently. While, in practice, it is not feasible to estimate and simulate multiple transition models for each status change or evaluate many assumptions, researchers should be more transparent about how these may affect outcomes and be wary that simple measures do not fully capture the uncertainty of the entire model.

For this analysis, only coefficient uncertainty and choices about model type, period, and complexity for one module of the simulation were considered. Even for this relatively simple setup, over half a million runs and many simulation hours were necessary. Indeed, many more possible choices could have been considered, e.g., regarding the time frame on which intercept adjustment values are based. Full analyses like these are computationally challenging, as run times for a single run can be several hours. Thus, effective ways of reducing computational load are necessary if sensitivity beyond the \ac{MC} effect and coefficient uncertainty are to be considered. One possibility would be the use of methods for experimental design to reduce the number of simulation runs.


\begin{acronym}[XXXXXX]
    \acro{BS}{Brier Score}
    \acro{BSS}{Brier Skill Score}
    \acro{MC}{Monte-Carlo}
    \acro{MNL}{Multinomial Logistic Regression}
    \acro{MSE}{Mean Squared Error}
    \acro{MZ}{Microcensus}
    \acro{RF}{Random Forest}
    \acro{SA}{Sensitivity Analysis}
    \acrodefplural{SA}{Sensitivity Analyses}
    \acro{SI}{Sensitivity Index}
    \acrodefplural{SI}{Sensitivity Indices}
    \acro{tSI}{total Sensitivity Index}
    \acrodefplural{tSI}{total Sensitivity Indices}
    \acro{mSI}{main-effect Sensitivity Index}
    \acrodefplural{mSI}{main-effect Sensitivity Indices}
    \acro{iSI}{interaction-effect Sensitivity Index}
    \acrodefplural{iSI}{interaction-effect Sensitivity Indices}
    \acro{SOEP}{Socio-Economic Panel}
    \acro{VC}{Variance Component}
\end{acronym}



\paragraph{Data and code availability statement} Raw data were generated at Trier University. The population data set is confidential due to legal constraints. Derived data supporting the findings of this study and codes are available from RM on request. 
\bibliography{literature}

\end{document}